\documentclass[floatfix,reprint,amsmath,amssymb,aps,prl,superscriptaddress,sort]{revtex4-2}
\def\selectlanguage#1{}

\usepackage{graphicx}
\usepackage{bm}
\usepackage{braket}
\usepackage{bbold}
\usepackage{tikz}
\usepackage{braids}
\usepackage{float}
\usepackage[caption=false]{subfig}
\usepackage[export]{adjustbox}
\usepackage[colorlinks=true, allcolors=blue]{hyperref}

\usetikzlibrary{arrows.meta, calc}
\DeclareMathOperator{\Tr}{Tr}

\begin{document}

\title{Interferometric braiding of non-Abelian anyons}

\author{Francesco Debortoli}
\email{f.debortoli@lmu.de}
\affiliation{Department of Physics and Arnold Sommerfeld Center for Theoretical Physics (ASC), Ludwig-Maximilians-Universität München, Theresienstr. 37, München D-80333, Germany}
\affiliation{Munich Center for Quantum Science and Technology (MCQST), Schellingstr. 4, München D-80799, Germany}
\affiliation{Max-Planck-Institute of Quantum Optics, Hans-Kopfermann-Str. 1, Garching D-85748, Germany}

\author{Fabian Grusdt}
\affiliation{Department of Physics and Arnold Sommerfeld Center for Theoretical Physics (ASC), Ludwig-Maximilians-Universität München, Theresienstr. 37, München D-80333, Germany}
\affiliation{Munich Center for Quantum Science and Technology (MCQST), Schellingstr. 4, München D-80799, Germany}

\date{\today}

\begin{abstract}
    The manipulation of non-Abelian anyons is a foundational ingredient for topological quantum computing and underlies topologically protected quantum memories. Here we analyze a quantum-impurity-based protocol for braiding non-Abelian anyons. Exploiting mobile impurities with addressable internal spin states, we provide a concrete protocol to experimentally measure the matrix elements of the $R$ and $F$ matrices that characterize the non-Abelian statistics of anyons. In particular, focusing on the case of Ising and Fibonacci anyons, we illustrate how the anyonic nature of the excitations emerges directly in the Ramsey interferometer we propose. Building on this framework, we further introduce a conventional-to-topological qubit encoder to transfer quantum information from conventional flying qubits into a topologically protected quantum memory.
\end{abstract}

\maketitle

\emph{Introduction. ---} Non-Abelian anyons offer degenerate many-body ground states that are particularly suitable for quantum computation, since they are naturally protected from decoherence. Braiding these quasiparticles allows one to perform quantum gates on the degenerate ground-state manifold \cite{nayak_non-abelian_2008,xu_non-abelian_2024,zhang_quasiparticle_2012}. So far, the exotic statistical properties of anyons have been experimentally probed mostly using edge-state interferometry \cite{nakamura_direct_2020}. At the same time, fractional Chern insulators realized in 2D materials have emerged as a new platform hosting topologically ordered ground states, often without the need for an external magnetic field \cite{wang_non-abelian_2012,bergholtz_topological_2013,liu_recent_2024,cai_signatures_2023,zeng_thermodynamic_2023,lu_fractional_2024,park_ferromagnetism_2025}.

These 2D platforms also host excitons --- charge-neutral quantum impurities that can be used to probe and optically control the material \cite{shimazaki_strongly_2020}. Recent theoretical \cite{mostaan_anyon-trions_2026,wagner_sensing_2026,lu_exciton-anyon_2026,wang_anyon-impurity_2026} and experimental \cite{li_signatures_2026} progress has further enabled the binding of mobile impurities to topological excitations in transition-metal dichalcogenides (TMDs). If such an impurity carries an internal (pseudo-)spin degree of freedom, a Ramsey interferometry protocol applied to the resulting bound state can measure the anyonic statistics of the topological excitations \cite{palm_interferometric_2026}. Still missing is a scheme for coherent manipulation and braiding of non-Abelian anyons directly in the bulk, complementary to established edge-state probes. Also missing is a hybrid interface able to encode (or teleport) quantum information from an unprotected, conventional qubit into a protected, topological one --- a powerful building block for quantum computation. To date, such interfaces have been realized only between conventional qubits, e.g., a flying photonic qubit and a semiconductor spin qubit \cite{gao_quantum_2013,gao_coherent_2015,mi_coherent_2018}.

\begin{figure}[b]
\centering
\vspace{-0.5cm}
\includegraphics[width=\columnwidth]{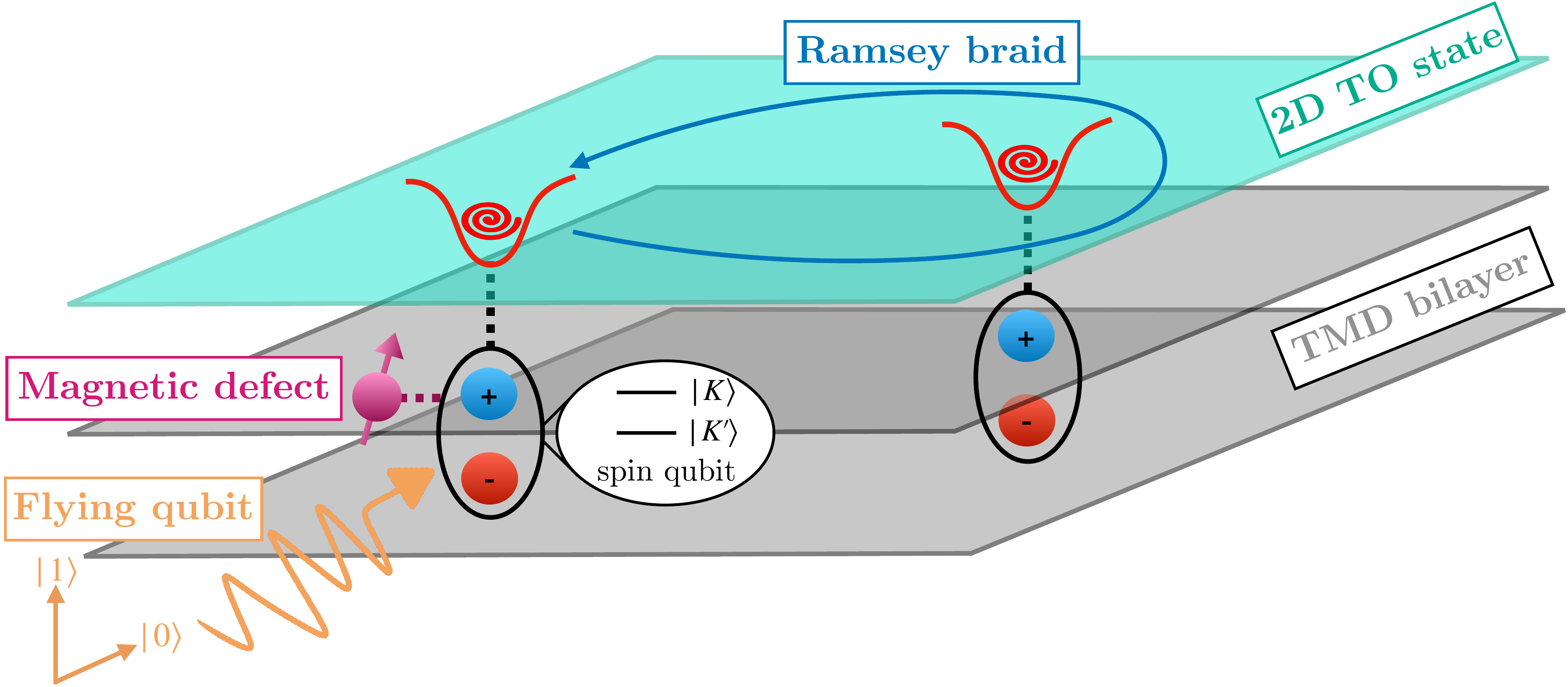}
\caption{Scheme of the proposed hybrid interface between the flying photonic qubit and the topological qubit. The topological qubit consists of excitations (represented as red vortices) on top of a 2D topologically ordered (TO) state. In a first step, the flying photonic qubit is used to initialize the spin qubit, encoded in the (pseudo-)spin state of the impurity. In a second step, the state of the spin qubit is encoded into the topological qubit using the braiding protocol detailed in the main text.}
\label{fig:setup}
\end{figure}

Here we propose a complementary approach, which works fully in the bulk of the system, opening a new route to probe non-Abelian statistics by braiding individual anyons in fractional Chern insulators or fractional quantum Hall states. Concretely, we envision interlayer excitons, realistically available in TMDs and possessing a valley degree of freedom, as the mobile impurity implementing this protocol, see Fig.~\ref{fig:setup}.

On the one hand, our protocol applies to arbitrary non-Abelian anyons and gives direct access to the $R$ and $F$ matrices that characterize their statistics \cite{nayak_non-abelian_2008}. We illustrate it for Ising and Fibonacci anyons, among the most relevant non-Abelian anyons for topological quantum computation, and show how measuring the resulting interference probability pattern allows one to directly identify their non-Abelian nature. On the other hand, the same setup constitutes a natural conventional-to-topological qubit interface: the internal state of the impurity, encoding a conventional qubit, can be coherently transferred into the topological one, thereby bridging conventional and topological quantum information processing.

\emph{The protocol. ---} We now extend the interferometric protocol proposed in Ref.~\cite{palm_interferometric_2026} to general non-Abelian anyons. We consider a generic topological state $\ket{\psi}$ with two quasiparticles localized at positions $\mathbf{r}_1$ and $\mathbf{r}_2$ (additional spectator quasiparticles may also be present, as in the multi-anyon examples discussed below). If these quasiparticles are non-Abelian anyons, braiding them transforms the system into a different degenerate ground state $\hat{W}\ket{\psi}$, where $\hat{W}$ is a unitary operator. The core idea is to bind each of the two quasiparticles to an impurity with two internal states, $\ket{\uparrow}$ and $\ket{\downarrow}$. In the following we assume that the impurity in the $\ket{\uparrow}$ state binds to the anyonic excitation and is itself moved during the sequence. The impurity in the $\ket{\downarrow}$ state also binds to the anyonic excitation but is not moved during the protocol. The Ramsey protocol \cite{palm_interferometric_2026} allows one to relate the expectation value $\braket{\psi|e^{-i\phi}\hat{W}+e^{i\phi}\hat{W}^\dagger|\psi}$ to a measurement of the probability of finding the two impurities in a given internal state. Here $\phi$ is a freely tunable control phase.

The measurement protocol, for an adiabatic exchange of two anyons trapped by impurities at positions $\mathbf{r}_1$ and $\mathbf{r}_2$, consists of the following steps. $(i)$ We start with two impurities in the $\ket{\downarrow}$ state, respectively bound to the anyons at positions $\mathbf{r}_1$ and $\mathbf{r}_2$. The initial state of the full system is given by $\ket{\downarrow, \mathbf{r}_1}\ket{\downarrow, \mathbf{r}_2}\ket{\psi}$. $(ii)$ We apply a $\frac{\pi}{2}$-pulse on both impurities, which sends the state $\ket{\uparrow} (\ket{\downarrow})$ to $\ket{+}\equiv\frac{\ket{\uparrow}+\ket{\downarrow}}{\sqrt{2}} \left(\ket{-} \equiv \frac{\ket{\uparrow}-\ket{\downarrow}}{\sqrt{2}}\right)$. $(iii)$ We move only the $\ket{\uparrow}$ component of each impurity --- together with the bound anyon --- from position $\mathbf{r}_1$ ($\mathbf{r}_2$) to position $\mathbf{r}_2$ ($\mathbf{r}_1$), leaving the $\ket{\downarrow}$ components trapped at their original positions. Together, this implements a braiding operation that entangles the impurities' spin state with the topological state, evolving the system into a superposition of $\ket{\psi}$ and $\hat{W}\ket{\psi}$. $(iv)$ We apply a controllable phase gate which sends the up state $\ket{\uparrow}$ to $e^{-\frac{i}{4}\phi}\ket{\uparrow}$ and the down state $\ket{\downarrow}$ to $e^{\frac{i}{4}\phi}\ket{\downarrow}$. $(v)$ We apply a $\frac{\pi}{2}$-pulse on both impurities. At the end of the protocol, the probability of measuring one impurity at position $\mathbf{r}_1$ and the other at $\mathbf{r}_2$, both in the $\ket{\uparrow}$ state, is given by
\begin{equation}
    \label{eq:probability_expression}
    p_{\uparrow\uparrow} = \frac{1}{8}\left(1+\frac{1}{2}\braket{\psi|e^{-i\phi}\hat{W}+e^{i\phi}\hat{W}^\dagger|\psi}\right),
\end{equation}
up to a dynamical phase $\Delta \phi_\text{dyn}$ that can be compensated by a shift of $\phi \rightarrow \phi - \Delta \phi_\text{dyn}/2$.

\begin{figure}[t]
\centering
\begin{tikzpicture}[scale=1.2, >=Stealth]
  % Superconductor plane
  \draw[fill=gray!10, draw=gray!30, rounded corners] (-0.5, -0.8) rectangle (4.5, 1.2);
  % Vortices as twirly symbols with labels
  \foreach \i/\x in {1/0.5, 2/1.5, 3/2.5, 4/3.5} {
    \node (V\i) at (\x, 0.2) [circle, inner sep=2pt] {};
    \node at (\x, -0.2) {$\sigma_{\i}$};
    % Twirly symbol (spiral)
    \draw[red!80!black, thick] (\x, 0.2) arc (180:0:0.03) arc (360:180:0.06) arc (180:0:0.09) arc (360:200:0.12);
  }

  % Braiding path for B_12 (purple)
  \draw[purple, thick, dashed, ->] (V1) .. controls (1.0, -0.4) .. (V2);
  \draw[purple, thick, dashed, ->] (V2) .. controls (1.0, 0.8) .. (V1);
  \node[purple, font=\small] at (1.0, 0.95) {$B_{12}$};
  
  % Braiding path for B_23 (blue)
  \draw[blue, thick, dashed, ->] (V2) .. controls (2, -0.4) .. (V3);
  \draw[blue, thick, dashed, ->] (V3) .. controls (2, 0.8) .. (V2);
  \node[blue, font=\small] at (2.0, 0.95) {$B_{23}$};
  
\end{tikzpicture}
\caption{Top-down view of four Ising anyons of type $\sigma$. The dashed blue (purple) lines represent the braiding of anyons $\sigma_2$ ($\sigma_1$) and $\sigma_3$ ($\sigma_2$). This operation corresponds to the braiding operator $\hat{B}_{23}$ ($\hat{B}_{12}$).}
\label{fig:vortex_layout}
\end{figure}
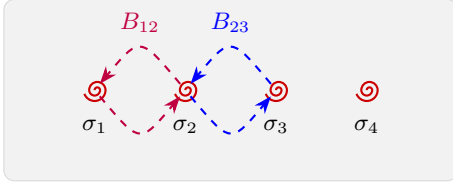

\emph{A possible experimental realization. ---}
A key ingredient to implement the protocol described in the previous section is the conditional displacement of the impurity depending on its internal (pseudo-)spin state. For its realization, we propose to use the valley degree of freedom of an interlayer exciton in a TMD bilayer as the internal pseudo-spin of the mobile impurity, $\ket{\uparrow}\equiv\ket{K}$ and $\ket{\downarrow}\equiv\ket{K'}$. The exciton binds to the anyon through the attraction between the (fractional) charge of the quasiparticle and the dipole moment of the exciton~\cite{mostaan_anyon-trions_2026,lu_exciton-anyon_2026}; this coupling carries no valley index, so both components have the same binding energy. To realize the state selectivity necessary for step $(iii)$ of the protocol, we envision that the $\ket{\downarrow}$ component of the excitons at positions $\mathbf{r}_1$ and $\mathbf{r}_2$ gets trapped by two localized magnetic defects, which generate a localized valley-dependent potential centered on the defects' positions (see Fig.~\ref{fig:setup}): an attractive well for $\ket{\downarrow}$ and a repulsive barrier for $\ket{\uparrow}$. Since the $\ket{\downarrow}$ component is trapped, only the $\ket{\uparrow}$ component is displaced in the subsequent steps of the protocol by a weak electric field coupling to the total charge of the exciton-anyon bound state. Further details on the use of magnetic defects to implement the braiding protocol are provided in the Supplemental Material.

\emph{Conventional-to-topological qubit encoder. ---} We show how the protocol described previously can be adapted to realize a conventional-to-topological qubit encoder. The main ingredient is a braiding operation that maps the original state of the topological qubit to a different, topologically degenerate ground state orthogonal to it. For definiteness, we consider a system of four Ising anyons of type $\sigma$ (see Fig.~\ref{fig:vortex_layout}). The Hilbert space with total topological charge $1$ is spanned by the two basis states $\ket{0}_\text{top}=\ket{(\sigma_1, \sigma_2)_1, (\sigma_3, \sigma_4)_1}$ and $\ket{1}_\text{top} = \ket{(\sigma_1,\sigma_2)_\psi, (\sigma_3,\sigma_4)_\psi}$. With the notation $(\sigma_i, \sigma_j)_\gamma$ (where $\gamma = 1, \psi$), we indicate that the anyons $\sigma_i$ and $\sigma_j$ are in the $\gamma$ fusion channel \cite{nayak_non-abelian_2008,ivanov_non-abelian_2001,leijnse_introduction_2012}. We assume that the topological qubit is initialized in the state $\ket{0}_{\text{top}}$. Moving the $\sigma_2$ anyon along a closed path around the $\sigma_3$ anyon maps the state to the orthogonal state $\ket{1}_\text{top}$ (with $\hat{B}_{23}^2\ket{0}_\text{top} = e^{-i \pi/4}\hat{\sigma}_x\ket{0}_\text{top}=e^{-i \pi/4}\ket{1}_\text{top}$). 

We propose a hybrid interface between a flying photonic qubit and the topological qubit, see Fig.~\ref{fig:setup}. In a first step, the flying photonic qubit is used to initialize the spin qubit in the desired state. Subsequently, the state of the impurity (the spin qubit) is encoded into the degenerate ground-state manifold of the four Ising anyons (the topological qubit) using the following protocol. As before, we assume that only the $\ket{\uparrow}$ component of the impurity is displaced. $(i)$ We start with an impurity localized at the same position as the $\sigma_2$ anyon. We call this position $\mathbf{r}_2$. The impurity is in the state $\alpha\ket{\downarrow, \mathbf{r}_2}+\beta\ket{\uparrow, \mathbf{r}_2}$. This represents our conventional qubit, with the complex coefficients $\alpha,\beta$ inherited from the state of the original flying qubit. $(ii)$ We apply a spin-dependent phase to the impurity state (for example via the optical AC Stark effect), which transforms the state of the system to $(\alpha \ket{\downarrow, \mathbf{r}_2} + e^{i\phi} \beta \ket{\uparrow, \mathbf{r}_2}) \ket{0}_\text{top}$. $(iii)$ The $\ket{\uparrow}$ component of the impurity is transported around the $\sigma_3$ anyon, which corresponds to the braiding operation $\hat{B}_{23}^2$. The final state of this operation is given by $\alpha \ket{\downarrow, \mathbf{r}_2} \ket{0}_\text{top} + \beta e^{i\phi} \ket{\uparrow, \mathbf{r}_2} \hat{B}_{23}^2 \ket{0}_\text{top} = \alpha \ket{\downarrow, \mathbf{r}_2} \ket{0}_\text{top} + \beta  e^{i(\phi-\pi/4)} \ket{\uparrow, \mathbf{r}_2}\ket{1}_\text{top}.$
Hereafter, we set $\phi = \pi/4$ for simplicity. $(iv)$ We apply a Hadamard gate to the spin qubit, which transforms the state of the system to $\frac{1}{\sqrt{2}}\ket{\downarrow, \mathbf{r}_2} (\alpha \ket{0}_\text{top} + \beta \ket{1}_\text{top}) + \frac{1}{\sqrt{2}}\ket{\uparrow, \mathbf{r}_2} (\alpha \ket{0}_\text{top} - \beta \ket{1}_\text{top}).$ $(v)$ We measure the spin state of the impurity at position $\mathbf{r}_2$. If we find the impurity in the state $\ket{\downarrow, \mathbf{r}_2}$, then the topological qubit is projected onto the state $\alpha \ket{0}_\text{top} + \beta \ket{1}_\text{top}$. If we find the impurity in the state $\ket{\uparrow, \mathbf{r}_2}$, then the topological qubit is projected onto the state $\alpha \ket{0}_\text{top} - \beta \ket{1}_\text{top}$, which is equivalent to the original spin-qubit state up to a Pauli $\hat{\sigma}_z$ operation.

\emph{Results for the interferometric pattern. ---} Next, we explicitly calculate the probability pattern (Ramsey fringes) resulting from the interferometric protocol introduced above for Ising and Fibonacci anyons. Ising anyons emerge in a variety of physical systems, such as the $\nu = 5/2$ fractional quantum Hall state \cite{moore_nonabelions_1991} and $p$-wave superconductors \cite{read_paired_2000,ivanov_non-abelian_2001}. Furthermore, the physics of Ising anyons is particularly relevant for cold atomic systems. Various proposals have been made to realize a $p$-wave superfluid of fermionic dipolar molecules \cite{fedorov_novel_2016,cooper_stable_2009}. We demonstrate how the predicted interference pattern strongly depends on the braiding operations performed before the interferometric protocol and on the anyon type. This pattern can thus be used as a hallmark of the non-Abelian nature of the excitations.

\begin{figure}[t]
\centering
\includegraphics[width=\columnwidth]{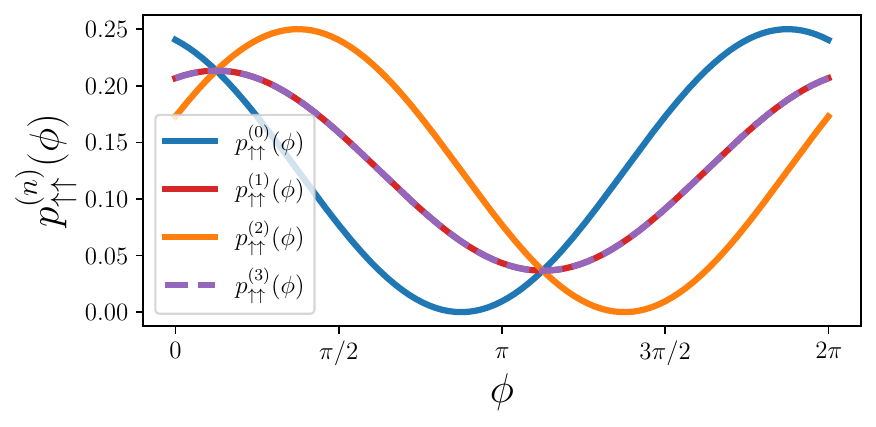}
\caption{Probabilities of finding both impurities in the $\uparrow$ state when the topological system is initialized in the state $\ket{0}_\text{top}$ for Ising anyons. The probability pattern goes to zero for $n=0,2$ braiding operations (see blue and orange curves), while it is always nonzero for $n=1,3$ braiding operations (see red and purple dashed curves) --- demonstrating the non-Abelian character of the anyons.}
\label{fig:pattern_ising}
\end{figure}

We start with the case of four Ising anyons (see Fig.~\ref{fig:vortex_layout}). Suppose that in the basis $\{ \ket{0}_\text{top}, \ket{1}_\text{top} \}$, the initial state of the topological system is described by the density matrix $\hat{\rho} = \begin{pmatrix} \rho_0 & \rho_{01} \\ \rho_{10} & \rho_1 \end{pmatrix}$. We then perform the braiding operation $\hat{B}_{23}$ $n$ times. The probability of finding both impurities in the up state at the end of the protocol is $p^{(n)}_{\uparrow\uparrow}(\phi) = \frac{1}{8} \left(1+\Tr((\hat{B}_{23}^n)^{\dagger} \hat{M}(\phi) \hat{B}_{23}^n \hat{\rho})\right)$, where $\hat{M}(\phi) = (e^{-i\phi} \hat{B}_{12} + e^{i\phi} \hat{B}_{12}^\dagger)/2$ is a Hermitian matrix and $\phi$ is a controllable phase. For Ising anyons, $\hat{B}_{12} = e^{i \alpha} e^{-i\frac{\pi}{4}\hat{\sigma}_z}\equiv\text{diag}(R_1^{\sigma \sigma}, R_\psi^{\sigma \sigma})= \text{diag}(e^{-i\pi/8}, e^{i3\pi/8})$, $\hat{B}_{23} = e^{i \alpha} e^{-i \frac{\pi}{4} \hat{\sigma}_x}$, and $\alpha = \pi/8$ is an Abelian phase. Therefore, $\hat{M}(\phi) = \begin{pmatrix}
    \cos(\phi+\pi/8) & 0 \\
    0 & \sin(\phi+\pi/8)
\end{pmatrix}$ and the probability patterns for $n=0,1,2,3$ braiding operations are given by $p^{(0)}_{\uparrow\uparrow}(\phi) = \frac{1}{8} \left(1 + \rho_0 \cos(\phi+\pi/8)+\rho_1 \sin(\phi+\pi/8)\right)$, $p^{(1)}_{\uparrow\uparrow}(\phi) = \frac{1}{8} (1 + \frac{1}{2}(\cos(\phi+\pi/8)+\sin(\phi+\pi/8)+i(\rho_{01} - \rho_{10})(\cos(\phi+\pi/8)-\sin(\phi+\pi/8))))$, $p^{(2)}_{\uparrow\uparrow}(\phi) = \frac{1}{8} \left(1+\rho_0\sin(\phi+\pi/8)+\rho_1\cos(\phi+\pi/8)\right)$, $p^{(3)}_{\uparrow\uparrow}(\phi) = \frac{1}{8} (1 + \frac{1}{2}(\cos(\phi+\pi/8)+\sin(\phi+\pi/8)-i(\rho_{01} - \rho_{10})(\cos(\phi+\pi/8)-\sin(\phi+\pi/8))))$.

The pattern is $4$-periodic in the number of braiding operations $n$, i.e. $p^{(n+4)}_{\uparrow\uparrow}(\phi) = p^{(n)}_{\uparrow\uparrow}(\phi)$. For simplicity, we specialize to the case $\rho_0 = 1$ and $\rho_1 = \rho_{01} = \rho_{10} = 0$. The probability patterns resulting from the interferometric protocol are shown in Fig.~\ref{fig:pattern_ising}. In this case, an experimental signature of the non-Abelian nature of the excitations emerges clearly. For $n=0$ braiding operations, the probability of finding both impurities in the up state goes to zero for $\phi = 7\pi/8$:
\begin{equation}
    p^{(0)}_{\uparrow\uparrow}(\phi) = \frac{1}{4} \cos^2\left(\frac{\phi+\pi/8}{2}\right) = \frac{1}{4} \cos^2\left(\frac{\phi-\arg R^{\sigma \sigma}_{1}}{2}\right).
\end{equation}
For $n=2$, the probability pattern is shifted by $\pi/2$ with respect to the $n=0$ case, and thus the probability of finding both impurities in the up state goes to zero for $\phi = 11\pi/8$. On the other hand, for $n=1$ and $n=3$ braiding operations, the probability of finding both impurities in the up state is always nonzero.

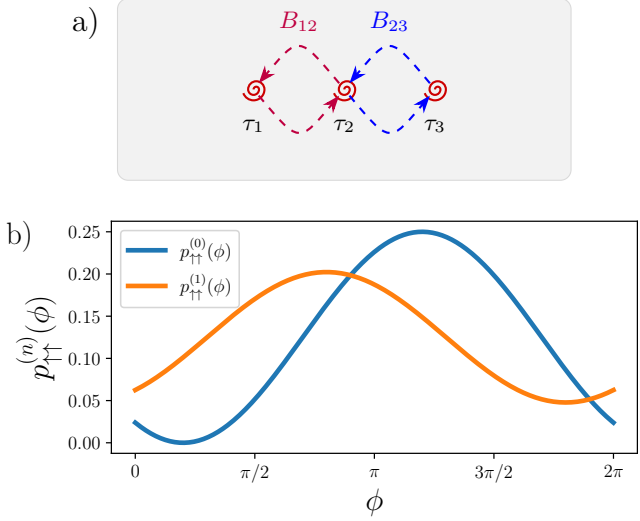
\begin{figure}[t]
\centering
\captionsetup[subfigure]{labelformat=empty}
\subfloat{\begin{tikzpicture}[scale=1.2, >=Stealth]
  % Superconductor plane
  \draw[fill=gray!10, draw=gray!30, rounded corners] (-0.5, -0.8) rectangle (4.5, 1.2);

  % Panel label
  \node[anchor=north east, font=\large] at (-0.6, 1.2) {a)};

  % Fibonacci anyons as twirly symbols with labels
  \foreach \i/\x in {1/1.0, 2/2.0, 3/3.0} {
    \node (V\i) at (\x, 0.2) [circle, inner sep=2pt] {};
    \node at (\x, -0.2) {$\tau_{\i}$};
    % Twirly symbol (spiral)
    \draw[red!80!black, thick] (\x, 0.2) arc (180:0:0.03) arc (360:180:0.06) arc (180:0:0.09) arc (360:200:0.12);
  }

  % Braiding path for B_12 (purple)
  \draw[purple, thick, dashed, ->] (V1) .. controls (1.5, -0.4) .. (V2);
  \draw[purple, thick, dashed, ->] (V2) .. controls (1.5, 0.8) .. (V1);
  \node[purple, font=\small] at (1.5, 0.95) {$B_{12}$};
  
  % Braiding path for B_23 (blue)
  \draw[blue, thick, dashed, ->] (V2) .. controls (2.5, -0.4) .. (V3);
  \draw[blue, thick, dashed, ->] (V3) .. controls (2.5, 0.8) .. (V2);
  \node[blue, font=\small] at (2.5, 0.95) {$B_{23}$};

\end{tikzpicture}\label{fig:fibonacci_braid}}

\subfloat{\includegraphics[width=\columnwidth, valign=c]{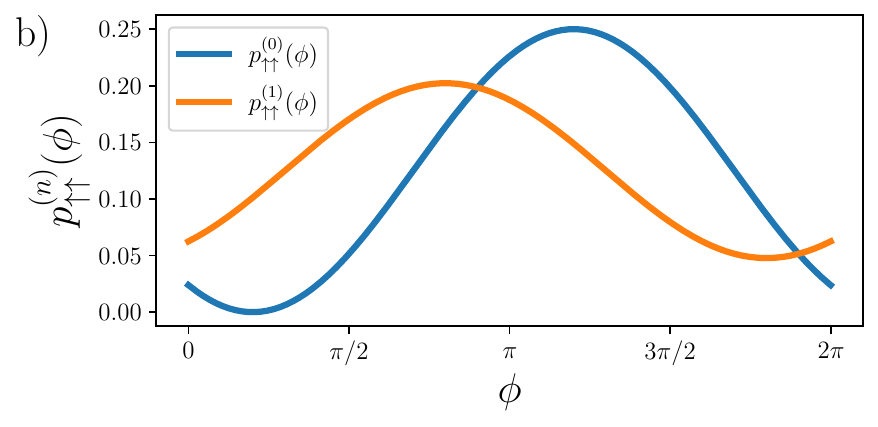}\label{fig:pattern_fibonacci}}
\vspace{-0.3cm}
\caption{\protect\subref{fig:fibonacci_braid} Top-down view of three Fibonacci anyons of type $\tau$. The dashed blue (purple) lines represent the braiding of anyons $\tau_2$ ($\tau_1$) and $\tau_3$ ($\tau_2$). This operation corresponds to the braiding operator $\hat{B}_{23}$ ($\hat{B}_{12}$). \protect\subref{fig:pattern_fibonacci} Probabilities of finding both impurities in the $\uparrow$ state when the topological system is initialized in the state $\ket{0}_\text{top}$ for Fibonacci anyons. The probability pattern goes to zero for $n=0$ braiding operations (blue curve), while it is always nonzero for $n=1$ braiding operations (orange curve) --- demonstrating the non-Abelian character of the anyons.}
\end{figure}

We now illustrate the interferometric protocol for Fibonacci anyons, showing that the probability pattern depends on the anyon type. A key feature of non-Abelian systems is the structure of their Hilbert space. In this case, to realize a qubit, we need three Fibonacci anyons (see Fig.~\ref{fig:fibonacci_braid}). A convenient basis for the Hilbert space of three Fibonacci anyons is given by $\{\ket{0}_\text{top} = \ket{((\tau_1, \tau_2)_1, \tau_3)_\tau}, \ket{1}_\text{top} = \ket{((\tau_1, \tau_2)_\tau, \tau_3)_\tau}, \ket{N} = \ket{((\tau_1, \tau_2)_\tau, \tau_3)_1} \}$. Braiding the first two quasiparticles $\tau_1$ and $\tau_2$ results in the braiding operator $\hat{B}_{12} = \text{diag}(R^{\tau \tau}_{1}, R^{\tau \tau}_{\tau}) \equiv \text{diag}(e^{-i 4\pi/5}, -e^{-i 2\pi/5})$, and the braiding of the second and third quasiparticles $\tau_2$ and $\tau_3$ results in the braiding operator
$\hat{B}_{23} = \begin{pmatrix}
    -e^{-\pi i/5}/\varphi & -i e^{-i\pi/10}/\sqrt{\varphi} \\
    -i e^{-i\pi/10}/\sqrt{\varphi} & -1/\varphi
\end{pmatrix}$, where $\varphi = (1+\sqrt{5})/2$ is the golden ratio. The matrix $\hat{M}(\phi)$ thus reads $\hat{M}(\phi) = \begin{pmatrix}
    \cos(\phi-6\pi/5) & 0 \\
    0 & \cos(\phi-3\pi/5)
\end{pmatrix}$. For Fibonacci anyons, the probability patterns after $n=0,1$ braiding operations are given by $p^{(0)}_{\uparrow\uparrow}(\phi) = \frac{1}{8} \left(1 + \rho_0 \cos(\phi-6\pi/5)+\rho_1 \cos(\phi-3\pi/5)\right)$ and $p^{(1)}_{\uparrow\uparrow}(\phi) = \frac{1}{8} (1-(\varphi^{-2}\rho_0+\varphi^{-1}\rho_1)\cos(\phi-\pi/5)+(\varphi^{-1}\rho_0+\varphi^{-2}\rho_1)\sin(\phi-\pi/10)+\varphi^{-\frac{3}{2}}(e^{i\frac{2\pi}{5}}\rho_{01}+e^{-i\frac{2\pi}{5}}\rho_{10})(\cos(\phi-\pi/5)+\sin(\phi-\pi/10)))$.

As in the Ising case, we specialize to the case $\rho_0 = 1$ and $\rho_1 = \rho_{01} = \rho_{10} = 0$. The probability patterns resulting from the interferometric protocol are shown in Fig.~\ref{fig:pattern_fibonacci}. In this case, the probability of finding both impurities in the up state without any braiding operation ($n=0$) goes to zero for $\phi = \pi/5$:
\begin{equation}
    p^{(0)}_{\uparrow\uparrow}(\phi) = \frac{1}{4} \cos^2\left(\frac{\phi-6\pi/5}{2}\right) = \frac{1}{4} \cos^2\left(\frac{\phi-\arg R^{\tau \tau}_{1}}{2}\right).
\end{equation}
After one braiding operation ($n=1$), the probability of finding both impurities in the up state is always nonzero.

This is a smoking-gun signature of non-Abelian statistics: for Abelian anyons, the Ramsey fringe always vanishes at $\phi$ equal to the exchange phase plus $\pi$. This is in direct analogy with the $n=0$ case discussed above, since Abelian anyons possess a single, well-defined fusion channel, so that braiding them is equivalent to braiding within a well-defined channel. We have thus shown that, while both Ising and Fibonacci anyons display this hallmark feature of non-Abelian statistics, the resulting fringe patterns differ markedly between the two models, reflecting their distinct fusion rules and braiding matrices.

\emph{Characterizing non-Abelian anyons: measuring the $R$ and $F$ matrices. ---} Non-Abelian anyons are characterized by the $R$ and $F$ matrices. The $R$ matrix encodes the phase resulting from the exchange of two anyons in a well-defined fusion channel. The $F$ matrix describes the change of basis between different pairings, or fusion trees, of the same anyons. We now detail how the interferometric protocol can be used to measure these matrices for general anyons, before discussing specifically the cases of Ising and Fibonacci anyons.

Consider three anyons of types $a$, $b$, and $c$ at positions $1$, $2$, and $3$, respectively (more anyons can be present, with $c$ representing their joint fusion outcome, as for the four-anyon Ising example below), with total topological charge $d$. Let $\ket{e}_{12} \equiv \ket{((a,b)_e, c)_d}$ denote the basis states in which the pair $(a,b)$ has well-defined fusion channel $e$. If we initialize the topological system in $\ket{\psi} = \ket{e}_{12}$ and bind the impurities to anyons $1$ and $2$, the interferometric protocol yields $p^{(0)}_{\uparrow\uparrow}(\phi) = \frac{1}{4} \cos^2\left((\phi-\arg R^{ab}_{e})/2\right)$, where we used the fact that $\hat{B}_{12}$ is diagonal in the $\ket{e}_{12}$ basis since $a$ and $b$ are in a well-defined fusion channel, i.e. $\braket{e|\hat{B}_{12}|e'} \equiv [\hat{B}_{12}]_{ee'} = R^{ab}_e \delta_{ee'}$. The resulting fringe pattern is thus a direct measurement of the $R$ matrix.

Repeating the protocol with the impurities instead bound to anyons $2$ and $3$, for a set of initial states $\ket{\psi}$, likewise allows the matrix elements $[\hat{B}_{23}]_{ee'}$ to be reconstructed. As we show below for Ising and Fibonacci anyons, the requirements on the initial-state preparation can often be relaxed depending on the anyon model --- e.g., it may suffice to prepare $\ket{0}_\text{top}$ and subsequently apply braiding operations.

The $F$ matrix relates $\ket{e}_{12}$ to the basis $\ket{g}_{23} \equiv \ket{(a,(b,c)_g)_d}$, in which $(b,c)$ instead has well-defined fusion channel $g$, via $\ket{e}_{12} = \sum_{g} [\hat{F}^{abc}_{d}]_{g e} \ket{g}_{23}$. Combined with the measured $R^{bc}_g$ and $[\hat{B}_{23}]_{ee'}$, the identity $[\hat{B}_{23}]_{e e'} = \sum_g [\hat{F}^{abc}_{d}]^{-1}_{e g}\,R^{bc}_g\,[\hat{F}^{abc}_{d}]_{g e'}$ then determines the $F$ matrix.

We now apply this general procedure to the anyon models introduced above, starting with Ising anyons: a topological system of four $\sigma$ anyons (see Fig.~\ref{fig:vortex_layout}), which we assume can be initialized in the logical state $\ket{0}_\text{top}$.

Binding the impurities to $\sigma_1$ and $\sigma_2$ and applying the protocol to $\ket{0}_\text{top}$ gives, via the general result above with $a=b=\sigma$ and $e=1$, $p_{\uparrow\uparrow}(\phi) = \frac{1}{4} \cos^2((\phi-\arg R^{\sigma \sigma}_{1})/2)$, measuring $R^{\sigma \sigma}_{1}$. Braiding $\sigma_2$ around $\sigma_3$ first transforms the initial state to $\hat{B}_{23}^2\ket{0}_\text{top} = e^{-i\pi/4}\ket{1}_\text{top}$, and repeating the protocol likewise gives $p_{\uparrow\uparrow}(\phi) = \frac{1}{4} \cos^2((\phi-\arg R^{\sigma \sigma}_{\psi})/2)$, reconstructing the full matrix $\hat{R} = \text{diag}(R^{\sigma \sigma}_{1}, R^{\sigma \sigma}_{\psi})$.

More generally, any braiding matrix $\hat{B}_{ij}$ can be reconstructed by operator tomography, realizing, for this two-level encoding, the reconstruction of $[\hat{B}_{23}]_{ee'}$ outlined above: preparing $\ket{\psi} = \hat{U} \ket{0}_\text{top}$ via an appropriate braiding operation $\hat{U}$ and binding the impurities to $\sigma_i, \sigma_j$ gives $\braket{\psi|\hat{M}(\phi)|\psi}$, with $\hat{M}(\phi) = (e^{-i\phi}\hat{B}_{ij}+e^{i\phi}\hat{B}_{ij}^\dagger)/2 \equiv M_0(\phi) \hat{\mathbb{1}}+M_x(\phi) \hat{\sigma}_x + M_y(\phi) \hat{\sigma}_y + M_z(\phi) \hat{\sigma}_z$ (see Eq.~\eqref{eq:probability_expression}). Measuring $p_{\uparrow\uparrow}$ for four choices of $\hat{U}$ determines $M_0, M_x, M_y, M_z$, and hence $\hat{B}_{ij} = \hat{M}(0)+i \hat{M}(\pi/2)$:

\begin{table}[H]
\centering
\begin{tabular}{|c|c|}
\hline
$\hat{U}$ (up to a phase) & $\braket{\psi|\hat{M}(\phi)|\psi}$\\ \hline
$\hat{\mathbb{1}}$ & $M_0(\phi) + M_z(\phi)$ \\ \hline
$\hat{B}_{23}^2 = \hat{\sigma}_x$ & $M_0(\phi) - M_z(\phi)$ \\ \hline
$\hat{B}_{23} = \exp\left(-i\frac{\pi}{4}\hat{\sigma}_x\right)$ & $M_0(\phi) - M_y(\phi)$ \\ \hline
$\hat{B}_{13} = \exp\left(-i\frac{\pi}{4}\hat{\sigma}_y\right)$ & $M_0(\phi) + M_x(\phi)$ \\ \hline
\end{tabular}
\end{table}
For Ising anyons, the only nontrivial $F$ matrix is $\hat{F}^{\sigma\sigma\sigma}_\sigma$, obtained from the general relation above, $\hat{B}_{23} = (\hat{F}^{\sigma\sigma\sigma}_\sigma)^{-1} \hat{R} \hat{F}^{\sigma\sigma\sigma}_\sigma$ --- an eigenvalue problem for its rows.

The same protocol applies to Fibonacci anyons, again assuming initialization in $\ket{0}_\text{top}$ and the ability to perform the $\pi/2$ rotations $\exp(-i\frac{\pi}{4} \hat{\sigma}_x), \exp(-i\frac{\pi}{4} \hat{\sigma}_y)$ on the topological qubit. Binding the impurities to $\tau_1, \tau_2$ measures $R^{\tau \tau}_{1}$ directly from $\ket{0}_\text{top}$ via $p_{\uparrow\uparrow}(\phi) = \frac{1}{4} \cos^2\left((\phi-\arg R^{\tau \tau}_{1})/2\right)$, and, after a $\hat{\sigma}_x$ gate prepares $\ket{1}_\text{top}$, the same measurement yields $R^{\tau \tau}_{\tau}$, reconstructing $\hat{R} = \text{diag}(R^{\tau \tau}_{1}, R^{\tau \tau}_{\tau})$. The matrices $\hat{B}_{ij}$ follow from the same tomographic protocol, and the only nontrivial $F$ matrix, $\hat{F}^{\tau\tau\tau}_\tau$, is obtained in analogy with the Ising case, $\hat{B}_{23} = (\hat{F}^{\tau\tau\tau}_\tau)^{-1} \hat{R} \hat{F}^{\tau\tau\tau}_\tau$. 

\emph{Conclusions. ---} In summary, we have demonstrated that impurity-based Ramsey interferometry can be effectively used to characterize the non-Abelian statistics of anyons by reconstructing the $R$ and $F$ matrices. We have provided explicit protocols for Ising and Fibonacci anyons, showing how the interferometric patterns can be used to unambiguously prove the non-Abelian character of the quasiparticles. We believe that interlayer excitons in TMD bilayers, together with magnetic defects supplying the valley-selective trapping, provide the most realistic platform for implementing the proposed protocol. Finally, we have introduced a conventional-to-topological qubit encoder, which establishes a concrete hybrid interface between the conventional qubit represented by the internal spin state of the impurity and the non-local topological qubit represented by topological excitations.

\emph{Acknowledgments. ---} The authors thank Nathan Goldman, Nader Mostaan, Felix Palm and Ata\c{c} \.{I}mamo\u{g}lu for insightful discussions. FD acknowledges support by the International Max Planck Research School. This work was funded by the Deutsche Forschungsgemeinschaft (DFG, German Research Foundation) under Germany's Excellence Strategy -- EXC-2111 -- 390814868, and by LMUexcellent, funded by the Federal Ministry of Education and Research (BMBF) and the Free State of Bavaria under the Excellence Strategy of the Federal Government and the Länder.

\bibliographystyle{apsrev4-2}
\bibliography{braiding.bib}

% ======================================================================
% Supplemental Material
% ======================================================================
\clearpage
\onecolumngrid
\begin{center}
\textbf{\large Supplemental Material for ``Interferometric braiding of non-Abelian anyons''}\\[0.5em]
\end{center}
\vspace{1em}
\twocolumngrid

\setcounter{equation}{0}
\setcounter{figure}{0}
\setcounter{table}{0}
\setcounter{section}{0}
\renewcommand{\thefigure}{S\arabic{figure}}
\renewcommand{\theequation}{S\arabic{equation}}
\renewcommand{\thetable}{S\arabic{table}}
\renewcommand{\thesection}{S\Roman{section}}

In this Supplemental Material we provide additional details on how to implement the proposed braiding protocol by making use of magnetic defects. As outlined in the main text, the key ingredient of the protocol is a state-dependent transport of the impurity. In particular, at step $(iii)$ of the protocol we want to move only the $\ket{\uparrow}$ components of the impurities, while leaving the $\ket{\downarrow}$ components at the initial positions $\mathbf{r}_1, \mathbf{r}_2$. We now detail how this can be achieved in practice, and discuss the relevant scales that control the protocol and explain why a defect breaking time-reversal symmetry is necessary to implement the required state selectivity.

In the following we consider a transition metal dichalcogenide (TMD) bilayer placed near a topologically ordered layer of interest, see Fig.~\ref{fig:setup}. In TMDs, interlayer excitons behave as tightly bound charge-neutral particles, with a permanent out-of-plane dipole moment $p=e\,d_z$. Throughout, we refer to the exciton binding energy as $E_X$.

The non-Abelian anyons hosted by the topologically ordered layer carry a (fractional) charge $e^{*}$ and interact with the exciton through the electrostatic interaction between the anyon charge and the exciton dipole. As recently demonstrated both theoretically \cite{mostaan_anyon-trions_2026,wagner_sensing_2026,lu_exciton-anyon_2026} and experimentally \cite{li_signatures_2026}, this interaction can give rise to a bound state between the anyon and the exciton, with a binding energy $E_b$ and a spatial extent $a$. Since this interaction is purely electrostatic, both valley components $\ket{K}, \ket{K'}$ of the exciton bind to the anyon with the same binding energy $E_b$. To understand this, note that time-reversal symmetry $\Theta$ maps the two valley excitons into each other, $\Theta\ket{K} = \ket{K'}$, and that Coulomb interactions are even under time-reversal.

Furthermore, the exciton-anyon bound state can be displaced using an in-plane electric field $\mathbf{E}(\mathbf{r}) \equiv -\bm{\nabla} U(\mathbf{r})$, which couples to the bound state's total charge $e^*$, resulting in a net force $\mathbf{F} = e^* \mathbf{E}$ acting on the bound state. Since this force is valley-independent, to implement step $(iii)$ of the protocol it needs to be supplemented by an additional ingredient sensitive to the valley degree of freedom, which we propose next.

We propose to trap the $\ket{K'} \equiv \ket{\downarrow}$ component of the excitons using two magnetic defects fixed at positions $\mathbf{r}_1, \mathbf{r}_2$. The magnetic defect breaks time-reversal symmetry, and therefore it can be used to implement a valley-dependent potential which acts as an attractive well for the $\ket{K'}$ component, while it acts as a repulsive barrier for the $\ket{K}$ component. In particular, the Hamiltonian of the exciton-anyon bound state in the presence of the magnetic defect positioned at $\mathbf{R}_d$ reads
\begin{equation}
    \hat H=\frac{\hat{\mathbf{p}}^{\,2}}{2M}
    +e^*U\!\left(\hat{\mathbf{r}}\right)\hat{\mathbb{1}}
    +\Delta\!\left(|\hat{\mathbf{r}}-\mathbf{R}_d|\right)\hat\sigma_{z},
\end{equation}
where $M$ is the bound state mass, $U$ is the valley-independent potential responsible for the bound state's displacement and $\Delta$ is the valley-dependent potential created by the magnetic defect. For definiteness, we consider a parabolic defect potential, $\Delta(r) = \Delta_0\left(1-\frac{r^2}{h^2} \right)$, where $\Delta_0>0$ is the depth of the well and $h$ is its range. To make sure that the $\ket{\downarrow}$ component remains trapped in the well, we require that the force exerted by the magnetic defect at a distance comparable to the bound state size $a$ is larger than the force exerted by the varying potential $U$: $\Delta_0 a/h^2 \gg |\mathbf{F}|$.

In summary, in this Supplemental Material we have discussed how interlayer excitons in TMD bilayers, together with magnetic defects providing a valley-dependent trapping, offer a realistic platform to implement the braiding protocol proposed in the main text.

\end{document}